\documentstyle{article}
\begin{document}
\title{Which Causality? \\Differences between Trajectory and
Copenhagen Analyses \\
of an Impulsive Perturbation \\}

\author{Edward R. Floyd \\
10 Jamaica Village Road, Coronado, California 92118-3208, USA \\
floyd@sd.cts.com \\}

\maketitle

\begin{abstract}

     The trajectory and Copenhagen representations render different
predictions for impulsive perturbations.  The different predictions
are due to the different roles that causality plays in the
trajectory and Copenhagen interpretations.  We investigate a small
perturbing impulse acting on the ground state of an infinitely deep
square well.  For the two representations, the first-order
perturbation calculations for the temporal change in energy differ. 
This temporal change in energy for the trajectory representation is
dependent upon the microstate of the wave function.  We show that
even under Copenhagen epistemology, the two representations predict
different theoretical results.\\

\end{abstract}



\noindent Suggested short running title:  Which Causality?

\noindent \scriptsize Type setting:  \LaTeX\ 2.09 with some \TeX. 
Author's files: PERTURB.TEX (input source code); PERTURB.DVI
(output)

\normalsize

\section{INTRODUCTION}

The trajectory interpretation has strong causality by precept. 
Bohr, the chief founder of the Copenhagen interpretation, held that
causality had to be dropped for a consistent quantum
theory.$^{\ref{bib:bohr}}$  Born, another founder of the Copenhagen
interpretation, asserted ``in the quantum theory it is the {\it
principle of causality}, or more accurately that of {\it
determination}, which must be dropped" (emphases in
original).$^{\ref{bib:born}}$ More recent champions of the
Copenhagen interpretation have construed ``causality" to be the
temporal evolution of the Schr\"{o}dinger wave function with
time.$^{\ref{bib:p&c},\ref{bib:schwabl}}$  (We shall use herein
this evolved usage of ``causality" with regard to the Copenhagen
interpretation.)  This causality for the Copenhagen interpretation
is a weak causality.  The different roles that causality plays in
the trajectory and Copenhagen interpretations are fundamental and
manifest different theoretical predictions.  

The Copenhagen interpretation of quantum mechanics has been
consistent with observations of large ensembles.  But it only
renders a probability of outcomes for individual events. 
Heretofore, the mainstream of physics has considered strong causal
theories of quantum mechanics to be only alternative
representations that rendered nothing new and, likewise, only
predicted probabilities for outcomes.  For example, Bohm's
stochastic causal theory of quantum mechanics, which purports to be
consistent with the probability amplitude of the wave function,
introduces chance by precept.$^{\ref{bib:b&h}}$ 

On the other hand, the trajectory representation is not based upon
chance.$^{\ref{bib:prd26}-\ref{bib:fpl9}}$  The trajectory itself
is strongly causal and deterministic.  The trajectory
representation has shown that the Born postulate of the Copenhagen
interpretation, which attributes a probability amplitude to the
Schr\"{o}dinger wave function, is unnecessary.$^{\ref{bib:afb20}}$
The trajectory representation has shown that the Schr\"{o}dinger
wave function of the Copenhagen interpretation is not an exhaustive
description of quantum phenomenon because the trajectory
representation distinguishes microstates, each specifying a
distinct trajectory, of the wave function.$^{\ref{bib:prd26}-
\ref{bib:fpl9}}$  Each trajectory or microstate is sufficient by
itself to specify the Schr\"{o}dinger wave
function.$^{\ref{bib:prd34}}$  We need not invoke an ensemble of
trajectories to get the Schr\"{o}dinger wave function.  The
trajectory representation renders deterministic predictions for an
individual particle viz-a-viz the probability predictions of the
Copenhagen representation for an ensemble of
particles.$^{\ref{bib:afb20}}$  

Herein, we investigate a situation where the type of causality
distinguishes the trajectory interpretation from the Copenhagen
interpretation.  We investigate a perturbing impulse acting on a
particle in the ground state of an infinitely deep square well. 
First-order perturbation calculations for the trajectory and
Copenhagen interpretations predict different results for the change
in energy.  A perturbing impulse acts on the particle wherever it
is in its trajectory at the instant of impulse.  In the Copenhagen
interpretation, the perturbing impulse acts upon the
Schr\"{o}dinger wave function as it is described at the instant of
impulse.  There exists a quantum operator, the Hamiltonian, for
making the Copenhagen prediction regarding energy.  

We recognize that the Copenhagen interpretation epistemologically
restricts in principle what can be measured.  We structure this
investigation to address those concerns of the Copenhagen school. 
While the trajectory is completely determined by a necessary and
sufficient set of initial
conditions,$^{\ref{bib:prd34},\ref{bib:prd29}}$ we may relax our
knowledge of initial conditions to satisfy Copenhagen epistemology
and still show different predictions.  This renders a counter
example where a strong causal theory, the trajectory
representation, can predict something new and different from the
Copenhagen interpretation even with Copenhagen epistemology.  An
investigation of the particle's energy during perturbation is
sufficient to establish this counter example.

We study a perturbing potential that has an impulse in time
described by a $\delta $-function and that is spatial symmetric,
linear and limited to a small but finite domain of the square well. 
This perturbing potential has been chosen for five reasons.  First,
it accentuates the difference in the temporal behavior of energy
for the two representations.  Second, first-order perturbation
theory suffices for exhibiting differences between the two
representations. Third, the physics of how this perturbation
temporally effects energy is easily understood.  Fourth, we can
spatially confine the perturbation to the neighborhood of the nodes
of the ground state (at the well's edge) of the infinitely deep
square well while we concurrently confine the perturbation to the
neighborhood of nodes of latent excited states.  This ensures that
all perturbation matrix elements will be small.  And fifth, it is
mathematically tractable as our computations use algebraic
expressions or elementary functions.  Where necessary,
approximations can be done by reasonable algebraic expressions that
do not lose any physics.  

We make it explicit that the trajectory representation considered
herein differs with pilot-wave representations.  They have
different equations of motion.  The trajectory representation does
not invoke a $\psi$-field.  The equations of motion for the
trajectory representation are determined by the Hamilton-Jacobi
transformation equation while Bohmian mechanics$^{\ref{bib:b&h}}$
assumes that the conjugate momentum should be the mechanical
momentum.  This leads to different results that have been discussed
elsewhere.$^{\ref{bib:prd26}}$  For a bound state of the square
well, Bohmian mechanics predicts that the bound particle always
stands still.$^{\ref{bib:pr85}}$  We also note that the
trajectories consider here are different from the trajectories for
the Feynman representation.  The Feynman representation uses a
classical generator of the motion as its
propagator$^{\ref{bib:f&h}}$ while the trajectory representation
uses a generalized Hamilton's characteristic function as its
generator of the motion.$^{\ref{bib:prd26}-\ref{bib:fpl9}}$   

The trajectory representation and the Schr\"{o}dinger
representation without the Copenhagen interpretation mutually imply
each other.$^{\ref{bib:fpl9}}$  The Born postulate of the
Copenhagen interpretation assigns a probability amplitude to the
Schr\"{o}dinger wave function.  Lest we forget, Schr\"{o}dinger
opposed the Copenhagen interpretation of his wave function.

Nothing herein implies that time-dependent perturbation theory for
a trajectory representation is easier than that for contemporary
wave mechanics.  It is much more difficult.  It is even more
difficult in general than classical canonical perturbation theory. 
But then, we ask so much more from trajectory perturbations.  

The rest of this paper is organized as follows.  Section 2
describes the trajectory representation of a particle in an
infinitely deep square well in the unperturbed state.  Section 3
addresses the differences between the time-dependent perturbation
of energy for the two representations.  Section 4 describes how the
investigation is structured to comply with Copenhagen epistemology. 
A brief discussion is given in Section 5 regarding our findings for
ensemble averages and any subsequent experimental verification.

\section{BEFORE PERTURBATION}   

\paragraph{Unperturbed System:}  Let us consider initially that a
particle is in the ground state of an infinitely deep square well
whose (unperturbed) potential is given by

\begin{equation}  
V(x) = \left\{ \begin{array}{cc}
            \infty, & \  \  |x|>q \\ [.08 in]
            0, &  \  \  |x|\leq q
            \end{array}
       \right.
\label{eq:sqv}
\end{equation}

\noindent where $q$ is finite positive.

The trajectory representation is based upon a generalized 
Hamilton-Jacobi formulation.  The time-independent generalized
Hamilton-Jacobi equation for quantum mechanics for a particle is
given for one-dimensional motion in the $x$-direction
by$^{\ref{bib:prd34}-\ref{bib:afb20},\ref{bib:pla214}}$ 
 
\begin{equation}
\frac{(\partial W/\partial x)^2}{2m}+V-E=-\frac{\hbar
^2}{4m}\langle W;x\rangle \label{eq:hje}
\end{equation}

\noindent where $\partial W/\partial x$ is conjugate momentum, $E$
is the energy for the particle, $m$ is mass of the particle, and
$\hbar=h/(2\pi)$ where in turn $h$ is Planck's constant.  The term
$\langle W;x\rangle $ in Eq.\ (\ref{eq:hje}) is the Schwarzian
derivative of Hamilton's characteristic function, $W$, with respect
to $x$.  The Schwarzian derivative is given by

\[
\langle W;x\rangle = \left[\frac{\partial ^3W/\partial
x^3}{\partial W/\partial x}-\frac{3}{2}\left(\frac{\partial
^2W/\partial x^2}{\partial W/\partial x}\right)^2 \right].
\]

\noindent The left side of Eq.\ (\ref{eq:hje}) manifests the
classical Hamilton-Jacobi equation while the Schwarzian derivative
on the right side manifests the higher order quantum effects.  The
general solution for $\partial W/\partial x$ is given
by$^{\ref{bib:prd34},\ref{bib:fpl9}}$ 

\begin{equation}
\partial W/\partial x = \pm (2m)^{1/2}(a\phi ^2+b\theta ^2+c\phi
\theta )^{-1}
\label{eq:cme}
\end{equation}

\noindent where $(a,b,c)$ is a set of real coefficients such that 
$a,b > 0$, and  $(\phi,\theta)$ is a set of normalized independent
solutions of the associated time-independent one-dimensional
Schr\"{o}dinger equation,

\[
\frac{-\hbar ^2}{2m} \frac{d^2\psi }{dx^2} + (V-E)\psi = 0 
\]

\noindent where $\psi $ is the Schr\"{o}dinger wave function.  The
independent solutions $(\phi,\theta)$ are normalized so that their
Wronskian, ${\cal W}(\phi,\theta) = \phi \, d\theta /dx - d\phi
/dx\, \theta $, is scaled to give  ${\cal W}^2(\phi,\theta) =
2m/[\hbar ^2(ab-c^2/4)] > 0$.  (The nonlinearity of the generalized
Hamilton-Jacobi equation induces this normalization upon ${\cal W}$
.$^{\ref{bib:fpl9}}$)  This ensures that $(a\phi ^2 + b\theta ^2 +
c\phi \theta) > 0$.  We note for completeness that a particular set
$(\phi,\theta)$ of independent solutions of the Schr\"{o}dinger
equation may be chosen by the superposition principle so that the
coefficient $c$ is zero.  

The motion in phase space is specified by Eq.\ (\ref{eq:cme}). 
This phase-space trajectory is a function of the set of
coefficients $(a,b,c)$.  The $\pm$ sign in Eq.\ (\ref{eq:cme})
designates that the motion may be in either $x$-direction.    
The corresponding solution for the generalized Hamilton's
characteristic function, $W$, is given
by$^{\ref{bib:prd34},\ref{bib:fpl9}}$ 

\begin{equation}
W=\hbar \arctan \left(\frac{b(\theta /\phi ) + c/2}{(ab-
c^2/4)^{1/2}}\right)+K
\label{eq:hcfe}
\end{equation}

\noindent where $K$ is an integration constant, which we may set to
zero herein.  

The unperturbed energy, $E$, and action variable, $J$, for the
ground state are quantized by $E=\hbar ^2 \pi
^2/(8mq^2)=J^2/(32mq^2)$.  The relationship between $E$ and $J$ is
consistent with classical mechanics where $E_{\mbox{\scriptsize
classical}}=J^2/(32mq^2)$.  The quantization of $E$ and $J$ are
independent of the coefficients $a,\ b$ and
$c$.$^{\ref{bib:prd34}}$  Subsequently, the wave number for the
ground state is specified to be $k=\pi /(2q)$.  

The set of independent solutions $(\phi ,\theta )$ for this
unperturbed square well is chosen such that $\phi $ represents the
symmetric bound state given by

\[
\phi  =  \left(\frac{2m}{\hbar ^2k^2(ab-c^2/4)}\right)^{1/4} \cdot
\left\{ \begin{array}{lr}
               {\displaystyle \lim_{{\cal V}\to \infty}} ({\cal
E}/{\cal V})^{1/2} \exp [-\kappa (x-q)] \to 0, & x>q \nonumber \\
[.08 in]
               \cos (kx), & -q \leq x \leq q \\  [.08 in]
               {\displaystyle \lim_{{\cal V}\to \infty}} ({\cal
E}/{\cal V})^{1/2} \exp [\kappa (x+q)] \to 0, & x<-q  \nonumber
               \end{array}
          \right.
\]

\noindent where ${\cal V} \to \infty $ manifests an infinitely deep
square well, where ${\cal E}$ is a function of ${\cal V}$ given
implicitly by the quantizing equation ${\cal E}^{1/2}=\hbar (2m)^{-
1/2}q^{-1} \arctan \{[({\cal V}-{\cal E})/{\cal E}]^{1/2}\}$ for
the finite square well, and where $\kappa$ is a function of ${\cal
V}$ given by $\kappa = [2m({\cal V}-{\cal E})]^{1/2}/\hbar$.  [Note
that $\lim_{{\cal V}\to \infty } {\cal E} \to \hbar ^2 \pi
^2/(8mq^2)=E$.]  The other solution, $\theta $, is unbound and is
not unique as any amount of $\phi $ may be added to it.  While
$\phi $ represents the symmetric bound state, the corresponding
$\theta $ that we have chosen is antisymmetric. We present this
unbound solution as

\[
\theta =  \left(\frac{2m}{\hbar ^2k^2(ab-c^2/4)}\right)^{1/4} \cdot
\left\{ \begin{array}{lr}
{\displaystyle \lim_{{\cal V}\to \infty }} \left( \frac{{\cal
V}}{{\cal V}-{\cal E}} \right)^{1/2} \left[ \cosh [\kappa (x-q)] -
\frac{{\cal E}}{{\cal V}} \exp [\kappa (x-q)] \right] \to \infty ,
& x>q \nonumber \\ [.08 in]
\sin (kx),  & -q \leq x \leq q \\ [.08 in]
{\displaystyle \lim_{{\cal V}\to \infty }} \left( \frac{{\cal
V}}{{\cal V}-{\cal E}} \right)^{1/2} \left[ -\cosh [\kappa (x+q)]
+ \frac{{\cal E}}{{\cal V}} \exp [\kappa (x+q)] \right] \to -
\infty , & x<-q. \nonumber \\
\end{array}
\right.
\]

\noindent We have retained the limiting process as an intermediate
step in describing $\phi $ and $\theta $ in the classically
forbidden region, $|x|>q$, to facilitate evaluating the Wronskian
there.  The corresponding Wronskian gives ${\cal W}^2(\phi ,\theta
) = 2m/[\hbar ^2(ab-c^2/4)] \geq 0$ everywhere as expected.  The
(unperturbed) ground state Schr\"{o}dinger wave function is well
known to be

\begin{equation}
\psi  =  \left\{ \begin{array}{lr}
               0, & x>q \nonumber \\ [.08 in]
               q^{-1/2} \cos (kx), & -q \leq x \leq q \\  [.08 in]
               0, & x<-q. \nonumber
               \end{array}
          \right.
\label{eq:gswf}
\end{equation}

\paragraph{Microstates:}  For bound states in general and the
ground state in particular, microstates of the Schr\"{o}dinger wave
function exist where the particular choice of the set of
coefficients $(a,b,c)$ specifies a unique trajectory in phase space
for a given quantized energy $E$ or quantized action variable $J$
as developed elsewhere.$^{\ref{bib:fpl9}}$  Each microstate is
consistent with the bound-state Schr\"{o}dinger wave function for
the bound state can also be expressed
by$^{\ref{bib:prd34},\ref{bib:fpl9}}$ 

\begin{eqnarray}
\psi & = & \frac{(2m)^{1/4} \cos(W/\hbar )}{[a-
c^2/(4b)]^{1/2}(\partial W/\partial x)^{1/2}} \nonumber \\
& = & \frac{(a\phi ^2+b\theta ^2+c\phi \theta )^{1/2}}{[a-
c^2/(4b)]^{1/2}} \cos \left[\arctan \left(\frac{b(\theta /\phi ) +
c/2}{(ab-c^2/4)^{1/2}}\right)\right] = \phi . 
\label{eq:tre}
\end{eqnarray}

\noindent  Thus, before the perturbing impulse acts on the
particle, each microstate of the ground state has the same
quantized energy as the Schr\"{o}dinger wave function.

Hamilton's characteristic function is a generator of motion.  The
equation of motion in the domain $[x,t]$ is given by the 
Hamilton-Jacobi transformation equation for a constant coordinate
(often called Jacobi's theorem).  The procedure simplifies for
coordinates whose conjugate momenta are separation constants. For
stationarity, $E$ is a separation constant for time.  Thus, the
equation of motion for the trajectory time, $t$, relative to its
constant coordinate $\tau$, is given as a function of $x$ by 

\begin{equation}
t-\tau= \partial W/\partial E
\label{eq:eom}
\end{equation}

\noindent where the trajectory for a given energy, $E$, is a
function of a set of coefficients $(a,b,c)$ and $\tau $ specifies
the epoch.

We see that a Hamilton-Jacobi development of the trajectory
representation does not return us to the classical representation
because we must consider microstates, which do not occur in an
infinitely deep, classical square well.  The trajectory associated
with a microstate of given energy E may be specified by the set of
coefficients $(a,b,c)$ or by the set of initial conditions
$(x_o,\dot{x}_o,\ddot{x}_o)$.$^{\ref{bib:prd34}}$  On the other
hand, classical motion in one dimension is simpler for it is
specified for a given $E$ by the initial condition $x_o$. 
Nevertheless, the particular microstate, $a=b$ and $c=0$, does
correspond to the contemporary concept of classical motion.

\section{DURING PERTURBATION}

\paragraph{Perturbing System:} Let us now apply a spatially
dependent impulse at time $\gamma $ to the quantum particle in the
ground state.  The impulse is spatially dependent and, where
occurring, is toward the center of the well.  The subsequent 
time-dependent perturbing potential, $F\Delta V(x,t)$, is a
spatially dependent impulse, which is symmetric, given by 

\begin{equation}
F\Delta V(x,t)=\delta (t-\gamma ) F \cdot \left\{ \begin{array}{lr}

          0, & |x|>q \mbox{  or  } |x|\leq q-\epsilon \nonumber \\
[.08 in]
          (-x-q+\epsilon ), & -q\leq x<-q+\epsilon  \\ [.08 in]
          (x-q+\epsilon ), &  q-\epsilon <x\leq q  \nonumber \\
          \end{array}
     \right.  
\label{eq:pv}
\end{equation}

\noindent where $F$ is a factor with units of force and where
$0<\epsilon \ll q$. The spatial dependence of $\Delta V$ is
intentionally symmetric to simplify our investigation.

Let us now calculate the first-order changes in energy induced by
$F\Delta V(x,t)$.  We compare the calculated changes both for a
Copenhagen interpretation based upon the standard time-dependent
perturbation theory for the Schr\"{o}dinger representation and for
a trajectory representation based upon a canonical, generalized
Hamilton-Jacobi perturbation theory.  Both theories are based upon
variation of parameters.  Different first-order predictions in $E$
are exhibited for the two perturbation methods.  First-order
changes in $E$ are sufficient for our comparison because we are
free to make $\epsilon $ and $F$ as small as necessary.

We first apply the standard wave-function time-dependent
perturbation theory.  The variation in the coefficient (parameter)
of the ground state, $C_0$ (the coefficient of the eigenfunction
that, by itself, specified the particle before time $\gamma $) is
well known and given by

\[
     i\hbar \dot{C}_0 = F(\Delta V)_{00}C_0 + O(F^2)
\]

\noindent where 

\[
(\Delta V)_{00}=\langle 0|\Delta V|0\rangle =\left[ \frac{\epsilon
^2}{2} - \frac{1}{4k^2} + \frac{\cos (2k\epsilon)}{4k^2} \right]
\]

Thus, $C_0$ varies to first order in $F$.  Therefore, as is well
known, the complex behavior of $C_0(t)$, to first order in $F$,
changes in angular frequency due to a finite $(\Delta V)_{00}$. 
Ergo, the first-order change in energy, $E_1$ of an ensemble of
particles to first order in $F$ is given by 

\begin{equation}
E_1 = F(\Delta V)_{00} = F \left[ \frac{\epsilon ^2}{2} -
\frac{1}{4k^2} + \frac{\cos (2k\epsilon)}{4k^2} \right].
\label{eq:v00}
\end{equation}

We now apply a canonical perturbation expansion, based upon a
generalized Hamilton-Jacobi theory to the trajectory
representation.  To do this, we allow the parameters $E$ and $\tau
$, which were constants of the motion in the unperturbed equation
of motion, Eq.\ (\ref{eq:eom}), to be varied as a function of time
by the perturbation.  (Energy is no longer a separation constant
for a time-dependent Hamiltonian; consequently, the ``constant
coordinate", $\tau $, is also no longer a constant coordinate.)  We
expand $E$ and $\tau $ in a power series in $F$.  As we are only
interested in the first order term of energy, we shall compute this
term only.
  
The generalized Hamilton-Jacobi equation, Eq.\ (\ref{eq:hje})
manifests the same canonical transformation that generates the
classical Hamilton-Jacobi transformation.$^{\ref{bib:gold}}$ 
Hence, the equations of motion for the variables $(E,\tau )$ for
the trajectory representation are the same as those for classical
perturbation theory.  These well known equations of motion are
given by 

\[
\dot{E}=F\partial \Delta V(E,\tau ,t)/\partial \tau \ \ \
\mbox{and} \ \ \ \dot{\tau }=-F\partial \Delta V(E,\tau
,t)/\partial E.
\]

\noindent If the perturbation is small [we are free to make
$(F,\epsilon )$ as small as required], then we can produce a 
first-order approximation $(E_1,\tau _1)$ to the time variation of
$(E,\tau )$ by using the constant, unperturbed values $(E_0,\tau
_0)$ on the right sides of the above equations.  For the
perturbation expansion of interest (the expansion of $E$), we must
solve

\begin{equation}
\dot{E_1}=-F\partial \Delta V(E_0,\tau _0,t)/\partial \tau _0.
\label{eq:cpe}
\end{equation}

\noindent In solving the above equations for our perturbation,
which is not an explicit function of momentum, we must express $x$
in $\Delta V$ as a function of $(E,\tau ,t)$.  This requires us to
turn the unperturbed equation of motion, Eq. (\ref{eq:eom}), inside
out.   

\paragraph{Reversion of $t(x)$ to $x(t)$:}  The unperturbed
equation of motion is given by the Hamilton-Jacobi transformation
equation, Eq.\ (\ref{eq:eom}), which renders

\begin{equation}
t-\tau =\frac{\pm 2 \frac{mx}{\hbar k}(ab-c^2/4)^{1/2}}{a+b +(a-b)
\cos (2kx) + c \sin (2kx)}
\label{eq:upeom}
\end{equation}

\noindent where the $\pm $ sign indicates that motion may be in
either direction.  The trajectory is dependent upon the
coefficients $(a,b,c)$, which specify the particular microstate. 
If $a\neq b$ or $c\neq 0$, then Eq.\ (\ref{eq:upeom}) is a
transcendental equation that cannot be turned inside out to solve
for $x$ in closed form. But the spatial dependence of $\Delta
V(x,t)$ is non-zero only over two small regions: $-q \leq x <-
q+\epsilon$ and $q-\epsilon <x \leq q$.  In these two regions we
may expand the trigonometric functions in a Maclaurin series about
$-q$ and $q$ respectfully.  For example, we may express Eq.\
(\ref{eq:upeom}) for a particle traveling in the $+x$-direction in
the domain $-q \leq x <-q+\epsilon$ as

\begin{equation}
t-\tau =\frac{\frac{mx}{\hbar k}(ab-c^2/4)^{1/2}}{b-ck(x+q)+(a-
b)k^2(x+q)^2}+O[(x+q)^3], \ \ 0\leq x+q\leq \epsilon \ll 1.
\label{eq:mseom}
\end{equation}

\noindent In Eq.\ (\ref{eq:mseom}), keeping the Maclaurin series
expansion of the trigonometric functions only up to second order
still captures the physical nuances of the microstates as specified
by the coefficients $(a,b,c)$ (again, $\epsilon $ may be made as
small as necessary).  If we ignore the $O[(x+q)^3]$ terms in Eq.\
(\ref{eq:mseom}), we may now turn the equation of motion inside out
to approximate $(x+q)$ as a quadratic solution to Eq.\
(\ref{eq:mseom}).  An approximate $(x+q)$ as a function of $(E,\tau
,t)$ may be given for a particle traveling in the $+x$-direction as

\begin{equation}
x+q\approx \frac{\left( \frac{c}{b}k+\frac{m}{\hbar kG(t-\tau
)}\right) - \left[ \left( \frac{c}{b}k+\frac{m}{\hbar kG(t-\tau
)}\right)^2 - 4\frac{a-b}{b} k^2 \left( 1 + \frac{mq}{\hbar kG(t-
\tau )} \right) \right] ^{1/2}}{2\frac{a-b}{b} k^2}, \ 0\leq
x+q\leq \epsilon \ll 1
\label{eq:ieom}
\end{equation}

\noindent where $G=b/(ab-c^2/4)^{1/2}$.  While we use $G$ to make
Eq.\ (\ref{eq:ieom}) less cumbersome, we note that $G$ is a
constant of the motion in its own right.  This follows for $G$ can
be expressed entirely by other constants of the motion and other
physical constants.  We have $G=(2m)^{1/2}I/(\hbar \cal W)$ where
$\cal W$ is, as previously given, the Wronskian for the set of
independent solutions $(\phi ,\theta )$ and $I$ is the Ermakov
invariant given by$^{\ref{bib:pla214}}$ $I=[a-c^2/(4b)]^{-1}$.  For
completeness, the constant of the motion $G$ by Eq.\
(\ref{eq:mseom}) can be related to the periodicity of the
microstate by  

\[
G=\frac{\hbar k}{m}\bigg/ \frac{4q}{t_{\mbox{\scriptsize period}}}
\]

\noindent where $t_{\mbox{\scriptsize period}}$ is the duration of
a single cycle in phase space of the microstate.

\paragraph{Perturbation Effects:}  We may now formulate $\Delta
V(E_0,\tau _0,t)$ for a particle traveling in the $+x$-direction in
the domain $-q \leq x <-q+\epsilon$ from Eqs.\ (\ref{eq:pv}) and
(\ref{eq:ieom}) as

\begin{equation}
\begin{array}{rl}
\Delta V \approx & - \frac{\delta (t-\gamma )}{2\frac{a-
b}{b}k^2}\left\{ \left( \frac{m}{\hbar kG(t-\tau
_0)}+\frac{c}{b}k\right) - \left[ \left( \frac{c}{b}k
+\frac{m}{\hbar kG(t-\tau _0)} \right)^2 -4 \frac{a-b}{b}k^2
\left(1 + \frac{m}{\hbar kG(t-\tau _0)}\right) \right] ^{1/2}
\right\}  \nonumber \\ [.3in]
      & \ \ \ \ -\delta (t-\gamma )\epsilon , \ \ \ \ \ \ \ \ -
\frac{mq}{\hbar kG} < \gamma - \tau _0\leq -\frac{m(q-
\epsilon)}{\hbar kG(1-\frac{c}{b}k\epsilon + \frac{a-
b}{b}k^2\epsilon ^2)}.
     \end{array}
\label{eq:ipv}
\end{equation}
   
\noindent where $E_0$ is manifested by $k$ for
$E_0=\hbar^2k^2/(2m)=\hbar ^2\pi ^2/(8mq^2)$. 

We now have the wherewithal to determine the right side of Eq.\
(\ref{eq:cpe}).  From Eqs.\ (\ref{eq:cpe}) and (\ref{eq:ipv}),
$\dot{E}_1$ is given by

\[
\begin{array}{rl} 
\dot{E}_1\approx  & \frac{\delta (t-\gamma )F}{2\frac{a-
b}{b}k^2}\left\{ \frac{-m}{\hbar kG(t-\tau _0)^2} +
\frac{\frac{m}{\hbar kG(t-\tau _0)^2}\left( \frac{c}{b}k+
\frac{m}{\hbar kG(t-\tau _0)}\right) -2\frac{a-b}{b}k^2
\frac{mq}{\hbar kG(t-\tau _0)^2}}{\left[
\left(\frac{c}{b}k+\frac{m}{\hbar kG(t-\tau _0)}\right)^2-4\frac{a-
b}{b}k^2\left( 1 + \frac{mq}{\hbar kG(t-\tau _0)}\right) \right]
^{1/2}} \right\}, \\ [.3in]
& \ \ \ \ \ \ \ \ \ \ \ \ \ \ \ \ \ \ \ \ \ \ \ \ \ \ \ \ \ \ \ -
\frac{mg}{\hbar kG} < \gamma -\tau _0 \leq -\frac{m(q-\epsilon
)}{\hbar kG(1-\frac{c}{b}k\epsilon + \frac{a-b}{b}k^2\epsilon ^2)}.
     \end{array}
\]

\noindent We may now integrate the above over the $\delta 
$-function duration to get

\begin{equation}
\begin{array}{rl}
E_1\approx  &  \frac{{\cal T}bF}{2(a-b)k^2}\left\{ \frac{-m}{\hbar
kG(\gamma -\tau _0)^2} + \frac{\frac{m}{\hbar kG(\gamma -\tau
_0)^2} \left(\frac{c}{b}k+\frac{m}{\hbar kG(\gamma -\tau
_0)}\right) -2\frac{a-b}{b}k^2
\frac{mq}{\hbar kG(\gamma -\tau _0)^2}}{\left[ \left(\frac{c}{b}k
+ \frac{m}{\hbar kG(\gamma -\tau _0)}\right)^2-4\frac{a-
b}{b}k^2\left( 1 + \frac{mq}{\hbar kG(\gamma -\tau _0)}\right)
\right] ^{1/2}} \right\}, \\ [.3in]
& \ \ \ \ \ \ \ \ \ \ \ \ \ \ \ \ \ \ \ \ \ \ \ \ \ \ \ \ \ \ \ \
\ -\frac{mg}{\hbar kG} < \gamma -\tau _0 \leq -\frac{m(q-\epsilon
)}{\hbar kG(1-\frac{c}{b}k\epsilon + \frac{a-b}{b}k^2\epsilon ^2)}
     \end{array}
\label{eq:e1e}
\end{equation}

\noindent where the ``$\cal T$" in ``${\cal T}bF$" in the above is
a unit measure in time.  This ``$\cal T$" manifests that $\delta 
$-function impulse has been integrated over a time domain that
includes the instant that the $\delta $-function acts.  If the
particle is between $-q$ and $-q+\epsilon $ at time $\gamma $, then
the particle and the system causing the perturbing force will
transfer energy between each other.  The amount of energy
transferred to first order in $F$ is given by the magnitude of
$E_1$.  The direction of transfer is given by the sign of $E_1$.
From Eq.\ (\ref{eq:e1e}), $E_1$ is a function of the particular
microstate as specified by the coefficients $(a,b,c)$.  The energy
for the particle is $E=E_0+E_1+O(F^2)$.  

Let us simplify $E_1$ by considering the particular microstate
where the coefficients are specified by $a=b$ and $c=0$.  Then,
Eq.\ (\ref{eq:ieom}) becomes exact, and Eq.\ (\ref{eq:e1e}) becomes
exactly

\begin{equation}
E_1\Big|_{a=b,c=0}=+F\frac{\hbar k}{m}{\cal T}, \ \ \ \-
\frac{mq}{\hbar k} < \gamma -\tau _0 \leq -\frac{m(q-\epsilon
)}{\hbar k}.
\label{eq:e1e0}
\end{equation}

\noindent This simplified case has an intuitive physical
interpretation.  We see from the plus sign on the right side of
Eq.\ (\ref{eq:e1e0}) that the perturbing system does work on the
particle by transferring to first order the energy $E_1$ to it. 
The amount of energy is proportional to the force factor, $F$, and
the distance that the particle transits against this force during
the perturbation's duration.  The transit distance for $a=b$ and
$c=0$ is given by the product of particle velocity, $\hbar k/m$,
and one unit of time denoted by ``$\cal T$" in Eq.\
(\ref{eq:e1e0}).  This is also what we would intuitively expect
from classical perturbation theory and is a manifestation that the
trajectory representation is strongly causal. 

We report that if the particle had been traveling in the 
$+x$-direction and had been located between $q-\epsilon $ and $q$
at time $\tau $, then $E_1$ would be given by

\begin{equation}
\begin{array}{rl}
E_1\approx  & \frac{{\cal T}bF}{2(a-b)k^2}\left\{ \frac{m}{\hbar
kG(\gamma -\tau _0)^2} - \frac{\frac{m}{\hbar kG(\gamma -\tau
_0)^2}\left( \frac{c}{b}k+\frac{m}{\hbar kG(\gamma -\tau
_0)}\right) -2\frac{a-b}{b}k^2
\frac{mq}{\hbar kG(\gamma -\tau _0)^2}}{\left[ \left(\frac{c}{b}k+
\frac{m}{\hbar kG(\gamma -\tau _0)}\right)^2-4\frac{a-
b}{b}k^2\left( 1 - \frac{mq}{\hbar kG(\gamma -\tau _0)}\right)
\right] ^{1/2}} \right\}, \\ [.3in]
& \ \ \ \ \ \ \ \ \ \ \ \ \ \ \ \ \ \ \ \ \ \ \ \ \ \ \ \ \ \ \ \
\ \ \frac{m(q-\epsilon )}{\hbar kG(1-\frac{c}{b}k\epsilon +
\frac{a-b}{b}k^2\epsilon ^2)} < \gamma - \tau <\frac{mq}{\hbar kG}.
     \end{array}
\label{eq:e2e}
\end{equation}

\noindent For $a=b$ and $c=0$, we have 

\[
E_1\Big|_{a=b,c=0}=-F\frac{\hbar k}{m}{\cal T}, \ \ \ \ \frac{m(q-
\epsilon )}{\hbar k} < \gamma -\tau _0 <\frac{mq}{\hbar k}.
\]

\noindent where the particle does work on the perturbing system by
transferring to first order the energy $E_1$ to it.

Now we consider the effect of the perturbation upon a particle
travelling in the $-x$-direction.  We arbitrarily assume that the
cycle of interest is one for which particle motion in the 
$+x$-direction occurs first.  When the particle reflects at the
well boundary $x=q$, the trajectory in phase space rounds a
singular point and jumps from the Riemann sheet for $+x$ motion for
that cycle to the Riemann sheet for $-x$ motion for that cycle. 
Innate to this jumping of Riemann sheets is a shift in the epoch
$\tau $.  As $t\big|_{x=q}$ must be the same for both Riemann
sheets, we have from Eq.\ (\ref{eq:eom}) that $\tau _- =\tau _+ +
2mq/(\hbar kG)$ where the subscript of $\tau $ denotes the
direction of motion.  Unless needed, the sign subscript for $\tau
$ will not be made explicit.

We report that if the particle had been traveling in the 
$-x$-direction and had been located between $q$ and $q-\epsilon $
at time $\gamma $, then $E_1$ would be given by

\begin{equation}
\begin{array}{rl}
E_1\approx & \frac{{\cal T}bF}{2(a-b)k^2}\left\{ \frac{-m}{\hbar
kG(\tau _0-\gamma )^2} + \frac{\frac{m}{\hbar kG(\tau_0 -\gamma
)^2}\left( \frac{c}{b}k+\frac{m}{\hbar kG(\tau _0-\gamma )}\right)
-2\frac{a-b}{b}k^2
\frac{mq}{\hbar kG(\tau _0-\gamma )^2}}{\left[ \left(\frac{c}{b}k+
\frac{m}{\hbar kG(\tau _0-\gamma )}\right)^2-4\frac{a-
b}{b}k^2\left( 1 - \frac{mq}{\hbar kG(\tau _0-\gamma )}\right)
\right] ^{1/2}} \right\}, \\ [.3in]
& \ \ \ \ \ \ \ \ \ \ \ \ \ \ \ \ \ \ \ \ \ \ \ \ \ \ \ \ \ \ \ \
\ \ \ \ \ \frac{mq}{\hbar kG} < \tau _0 -\gamma  \leq \frac{m(q-
\epsilon )}{\hbar kG(1-\frac{c}{b}k\epsilon + \frac{a-
b}{b}k^2\epsilon ^2)}.
     \end{array}
\label{eq:e3e}
\end{equation}

\noindent For $a=b$ and $c=0$, we now have 

\[
E_1\Big|_{a=b,c=0}=+F\frac{\hbar k}{m}{\cal T}, \ \ \ \ \ \
\frac{m(q-\epsilon )}{\hbar k} < \gamma -\tau _0 <\frac{mq}{\hbar
k}.
\]

\noindent For motion in the $-x$-direction, we see that the
perturbing system does work on the particle as expected.

We report that if the particle had been traveling in the 
$-x$-direction and had been located between $-q+\epsilon $ and $-q$
at time $\gamma $, then $E_1$ would be given by

\begin{equation}
\begin{array}{rl}
E_1\approx & \frac{{\cal T}bF}{2(a-b)k^2}\left\{ \frac{m}{\hbar
kG(\tau _0-\gamma )^2} - \frac{\frac{m}{\hbar kG(\tau_0 -\gamma
)^2}\left( \frac{c}{b}k + \frac{m}{\hbar kG(\tau _0-\gamma
)}\right) -2\frac{a-b}{b}k^2
\frac{mq}{\hbar kG(\tau _0-\gamma )^2}}{\left[ \left(\frac{c}{b}k+
\frac{m}{\hbar kG(\tau _0-\gamma )}\right)^2-4\frac{a-
b}{b}k^2\left( 1 + \frac{mq}{\hbar kG(\tau _0-\gamma )}\right)
\right] ^{1/2}} \right\}, \\ [.3in]
& \ \ \ \ \ \ \ \ \ \ \ \ \ \ \ \ \ \ \ \ \ \ \ \ \ \ \ \ \ \ \ \
\ \ \frac{-m(q-\epsilon )}{\hbar kG(1-\frac{c}{b}k\epsilon +
\frac{a-b}{b}k^2\epsilon ^2)} \leq \tau _0 -\gamma  <\frac{-
mq}{\hbar kG}.
     \end{array}
\label{eq:e4e}
\end{equation}

\noindent For $a=b$ and $c=0$, we again have 

\[
E_1\Big|_{a=b,c=0}=-F\frac{\hbar k}{m}{\cal T}, \ \ \ \ \ \ \
\frac{m(q-\epsilon )}{\hbar k} < \gamma -\tau _0 <\frac{mq}{\hbar
k}
\]

\noindent as expected.  

If the particle is located between $-q+\epsilon $ and $q-\epsilon
$ at time $\gamma $, then 

\begin{equation}
E_1=0, \ \ \ \ \ \  |\tau - \gamma | \leq \left| \frac{m(q-\epsilon
)}{\hbar kG(1-\frac{c}{b}k\epsilon + \frac{a-b}{b}k^2\epsilon
^2)}\right|.
\label{eq:e5e}
\end{equation}

\noindent But we note that the first order $E_1$ is zero here for
any microstate of the trajectory representation because the
particle is not located in a space-time location where the
perturbation acts. 

Under the trajectory interpretation, we do know where the particle
is in trajectory theory: the trajectory representation is strongly
causal.  The time dependence of the perturbing system is a $\delta
$-function.  Ergo, we can predict individual results, $E_1$, in
trajectory theory by Eqs.\ (\ref{eq:e1e}) and (\ref{eq:e2e})--
(\ref{eq:e5e}).  These trajectory predictions for $E_1$ differ with
the predictions, Eq.\ (\ref{eq:v00}), of the Copenhagen
interpretation that are based upon ensemble averages!  Furthermore,
the trajectory predictions are a function of the particular
microstate.

\section{COPENHAGEN EPISTEMOLOGY}

Copenhagen epistemology rejects in principle sufficient knowledge
of the initial conditions to specify an individual trajectory
(microstate).  The Copenhagen school denies strong causality for
describing a particle's progress along a trajectory but rather
limits causality to the evolution of $\psi $.  Fortunately,
complete knowledge of the initial conditions attendant with strong
causality, although sufficient, is not necessary to show
differences.  We are free to apply a trajectory analysis with
incomplete information.  For the sake of this investigation, we
shall assume that we do not know the necessary and sufficient
initial conditions and that we must use ensemble averages.  Even
with ensemble averages, we may still show that the trajectory and
Copenhagen interpretations predict different first-order
perturbation energies.   

Had we not known the particle's position at time $\gamma $ due to
some practical limitation and, consistent with the trajectory
interpretation, not due to some limitation in principle, then if we
assume a uniform distribution of $\tau $ over the duration of one
cycle, the average $E_1$ as determined by Eqs.\ 
(\ref{eq:e1e}) and (\ref{eq:e2e})--(\ref{eq:e5e}), would be given
by $\langle E_1\rangle _{\mbox{\scriptsize average}}=0$.  Lest we
forget, each microstate (trajectory) by itself is sufficient to
deduce $\psi $.  Physically, averaging $E_1$ for a uniform
distribution of $\tau $ manifests that, to first order, the
perturbing system is as likely to do work on the particle as the
particle is likely to do work on the perturbing system.  We do not
need to know the microstate of any particular particle of the
ensemble.  It is sufficient that the ensemble be large enough so
that the sub-ensemble for each microstate would tend, with
increasing sub-ensemble size, toward a relative balance between
particles going in each direction along the $x$-axis.  This
relative balance is so because each microstate's orbit in phase
space has mirror symmetry about the configuration $x$-axis in
accordance with Eq.\ (\ref{eq:cme}).  Hence, we may generalize our
findings to any distribution of microstates so long as the 
sub-ensemble of each microstate is sufficiently large.  In such
case, we still have that $\langle E_1 \rangle _{\mbox{\scriptsize
average}}=0$ for the trajectory interpretation while the Copenhagen
interpretation finds that $E_1$ is finite as given by Eq.\
(\ref{eq:v00}).

One is cautioned not to infer from $\langle E_1\rangle
_{\mbox{\scriptsize average}}=0$ in the preceding paragraph for a
uniform distribution of $\tau $ that the expectation value of $E_1$
in the trajectory representation would be given to first order by
the time average of the perturbing potential evaluated in the
unperturbed state.  For perturbations of long duration, $E_1$ in
contemporary quantum mechanics is given to first order by the
expectation value of the perturbing potential in the unperturbed
state, which in classical mechanics corresponds to the time average
of the perturbation evaluated in the unperturbed
system.$^{\ref{bib:park}}$  But we have here an impulsive
perturbation.  Even for classical mechanics, $\langle E_1\rangle
_{\mbox{\scriptsize average}}$, as generated by an impulsive
perturbation with a uniform distribution of $\tau $, would be zero
and is not equivalent to the time average of the perturbing
potential evaluated in the unperturbed system.  Thus, classical
mechanics offers a precedent for impulsive perturbation.

Usually, the orbital period is assumed to be shorter than the
duration of the perturbation in classical systems.  As noted in the
preceding paragraph, this assumptions would allow us to simplify by
averaging over the orbit.  But we consider here the reverse
situation with an impulsive perturbation.  Immaterial of how much
we know, what happens by strong causality is determined by the
particle's position at the time of the perturbing impulse and what
it is doing then [i.e., $x(\gamma ), \dot{x}(\gamma )$ and
$\ddot{x}(\gamma )$] as described by a particular microstate of the
unperturbed energy.  These initial conditions are more extensive
then those needed for classical perturbation theory.  In classical
mechanics, the unperturbed energy, the applicable Riemann sheet of
$W_{\mbox{\scriptsize classical}}$ and the particle's position,
$x(\gamma )$, at the time of perturbing impulse suffice for
classical causality.

\section{DISCUSSION}

The predictions for an ensemble average $E_1$ of the trajectory and
Copenhagen interpretations differ.  In this ensemble, we need not
know the distribution of microstates nor the exact distribution of
the constant coordinate, $\tau $.  Hence, this comparison of
ensemble averages between the trajectory and Copenhagen
interpretation complies with Copenhagen epistemology.  The
underlying reason for the difference in the predictions between the
Copenhagen interpretation and the trajectory representation is due
to their difference regarding causality.  The trajectory
interpretation is strongly causal for individual trajectories while
the Copenhagen interpretation is weakly causal for ensembles while
only giving probabilities for individual events.  Any subsequent
experimental verification of our results is actually a test of
which causality quantum mechanics observes.

The particular problem that we have examined here was chosen to
simplify the mathematics in our exposition.  We need not be
constrained to such ideal conditions in any experimental
verification.  We do advise that the duration of any perturbation
be much shorter than the orbital period of the test particle.  (Had
we examined a small perturbation of long duration relative to the
particles orbital period, then the perturbing potential would
produce a change in energy given by the time average of the
perturbing potential evaluated by the unperturbed system which
would correspond to the expected first-order change in energy in
wave mechanics.$^{\ref{bib:park}}$)  A perturbation of very short
duration with spatial dependence would generate predictions that
accentuate the different roles that causality plays in the
trajectory and Copenhagen interpretations.

\clearpage

\paragraph{References}
\begin{enumerate}\itemsep -.06in
\item \label{bib:bohr} N. Bohr, {\it Atomic Theory and the
Description of Nature} (Cambridge University Press, Cambridge 1934)
pp.\ 48--54, 108.
\item \label{bib:born} M. Born, {\it The Restless Universe} (Dover,
New York, 1951) p.\ 51. 
\item \label{bib:p&c} J. L. Powell and B. Crasemann, {\it Quantum
Mechanics} (Addison-Wesley, Reading, MA.\ 1961) pp.\ 82--83.  
\item \label{bib:schwabl} F. Schwabl, {\it Quantum Mechanics}, 2nd
ed.\ (Springer-Verlag, New York, 1995) pp.\ 363--365.. 
\item \label{bib:b&h} D. Bohm and B. J. Hiley, {\it Phys.\ Rep.}
{\bf 144}, 323 (1987).
\item \label{bib:prd26} E. R. Floyd, {\it Phys.\ Rev.} {\bf D 26},
1339 (1982).
\item \label{bib:prd34} E. R. Floyd, {\it Phys.\ Rev.} {\bf D 34},
3246 (1986).
\item \label{bib:fpl9} E. R. Floyd, {\it Found.\ Phys.\ Lett.} {\bf
9}, 489 (1996).
\item \label{bib:afb20} E. R. Floyd, {\it An.\ Fond.\ Louis de
Broglie} {\bf 20}, 263 (1995).
\item \label{bib:prd29} E. R. Floyd, {\it Phys.\ Rev.} {\bf D 29},
1842 (1984).
\item \label{bib:pr85} D. Bohm, {\it Phys. Rev.} {\bf 85}, 166
(1952).
\item \label{bib:f&h} R. P. Feynman and A. R. Hibbs, {\it Quantum
Mechanics and Path Integrals} (McGraw-Hill, New York, 1965) pp.\
26--28.
\item \label{bib:pla214} E. R. Floyd, {\it Phys.\ Lett.} {\bf A
214}, 259 (1996).
\item \label{bib:gold} H. Goldstein, {\it Classical Mechanics}, 2nd
ed.\ (Addison-Wesley, Reading, MA. 1980) pp.\ 438--442.
\item \label{bib:park} D. Park, {\it Classical Dynamics and Its
Quantum Mechanical Analogues}, 2nd ed. (Springer-Verlag, New York,
1990) pp.\ 198, 232. 
\end{enumerate}

\clearpage

\begin{center}

{\LARGE ERRATA:  WHICH CAUSALITY?  DIFFFERENCES\\
BETWEEN TRAJECTORY AND COPENHAGEN ANALYSES\\
 OF AN IMPULSIVE PERTURBATION,\\
{\it Int. J. Mod. Phys.} {\bf A 14}, 1111 (1999)}

\bigskip

{Edward R. Floyd\\
10 Jamaica Village Road, Coronado, CA 92118-3208, USA\\
floyd@crash.cts.com}

\bigskip

Errata published: {\it Int.\ J.\ Mod.\ Phys.}\ {\bf A 16} (2001) 2447

\bigskip

\begin{minipage}{5.in}
{\small {\bf Abstract:} \ \ The evaluation of the matrix element $(\Delta V)_{00}$ is corrected.  This correction 
increases the differences between the trajectory and Copenhagen analyses.}
\end{minipage}

\end{center}

\bigskip

The unnumbered displayed equation preceding Eq. (9) is corrected by including the 
factor $\delta (t-\tau )/q$ so that $(\Delta V)_{00}$ becomes$^1$

\[
(\Delta V)_{00} =  \langle 0|\Delta V|0 \rangle = \frac{\delta (t-\gamma )}{q} \left[ 
\frac{\epsilon  ^2}{2} - \frac{1}{4k^2} + \frac{\cos(2k\epsilon )}{4k^2} \right]
\]

\noindent Consequently, the correct first-order change in energy, $E_1$, of an 
ensemble of particles is given by correcting Eq. (9) to read 

\setcounter{equation}{8}

\begin{equation}
E_1 = F(\Delta V)_{00} =  \frac{F\delta (t-\gamma  )}{q} \left[ \frac{\epsilon  
^2}{2} - \frac{1}{4k^2} + \frac{\cos(2k\epsilon )}{4k^2} \right].
\end{equation}

Including this correcting factor $\delta (t-\tau )/q$ increases the magnitude of 
$E_1$ during perturbation for the Copenhagen interpretation.  Meanwhile, the 
average $E_1$ of the ensemble of particles for the trajectory representation under 
Copenhagen epistemology remains $\langle E_1 \rangle _{\mbox {average}} = 0$ 
during perturbation.$^1$  This strengthens our findings that trajectory and 
Copenhagen analyses differ regarding perturbing impulses. 

\paragraph{References}

\begin{enumerate}
\item E. R. Floyd, {\it Int. J. Mod. Phys.} {\bf A 14}, 1111 (2000).
\end{enumerate}

\end{document}